\documentclass[referee]{aastex63}

\usepackage{xcolor}

\usepackage{bm}
\usepackage{amsmath}
\usepackage{empheq}
\usepackage{enumitem}

\received{XXX}
\revised{XXX}
\accepted{\today}

\def\tff{\tau_{\rm ff,0}}
\def\tg{\tau_{\rm G,0}}
\def\tmean{\bar{\tau}}
\def\tcoll{t_{\rm coll}}
\def\vir{\alpha_{\rm vir}}

\DeclareUnicodeCharacter{00A0}{ }

\shorttitle{Density PDFs}
\shortauthors{Jaupart \& Chabrier}
\graphicspath{{./}{figures/}}

\begin{document}

\title{Evolution of the density PDF in star forming clouds: the role of gravity.}



\author{Etienne Jaupart}
\affiliation{Ecole normale sup\'erieure de Lyon, CRAL, Universit\'e de Lyon, UMR CNRS 5574, F-69364 Lyon Cedex 07, France}

\author{Gilles Chabrier}
\affiliation{Ecole normale sup\'erieure de Lyon, CRAL, Universit\'e de  Lyon, UMR CNRS 5574, F-69364 Lyon Cedex 07, France}
\affiliation{School of Physics, University of Exeter, Exeter, EX4 4QL, UK}

\email{etienne.jaupart@ens-lyon.fr, chabrier@ens-lyon.fr}

\begin{abstract}

We  derive an analytical theory of the PDF of density fluctuations in supersonic turbulence in the presence of gravity in star-forming clouds. The theory is based on a rigorous
derivation of a combination of the Navier-Stokes continuity equations for the fluid motions and the Poisson equation for the gravity. It extends upon previous approaches first by including gravity, second by considering the PDF as a dynamical system, not a stationary one. We derive  the  transport equations  of 
the density PDF, characterize its  evolution and determine the density threshold above which gravity strongly affects and eventually dominates the dynamics of  turbulence.  
We demonstrate the occurence of {\it two} power law tails in the PDF, with two characteristic exponents, corresponding to two different stages in the balance between turbulence and gravity. 
Another important result of this study is to provide a procedure to relate the observed {\it column density} PDFs to the corresponding {\it volume density} PDFs. This allows to infer, from the observation of column-densities, various physical parameters characterizing molecular clouds, notably the virial parameter. 
Furthermore, the theory offers the possibility to date the clouds in
units of ${t}_{\rm coll}$, the time  since a statistically significant fraction of the cloud started to collapse. The theoretical results and diagnostics reproduce very well  numerical simulations and observations of star-forming clouds. The theory provides a sound theoretical foundation and quantitative diagnostics to analyze observations or numerical simulations of star-forming regions and to characterize the evolution of the density PDF in various regions of molecular clouds.

\end{abstract}



\section{Introduction} \label{sec:intro}

It has been established by many studies that the volume-weighted Probability Density Function (PDF) of supersonic isothermal turbulence displays a nearly lognormal shape for solenoidally driven turbulence, at least for Mach numbers ${\cal M}\lesssim 30$ (\citealt{vazquez1994, passot1998, kritsuk2007,federrath2008,federrath2010,Pan2019B}) even in the presence of magnetic fields (\citealt{lemaster2008, collins2012}). In dense star-forming regions, however, the line-of-sight extinction and inferred column density PDFs have been observed to develop a power-law tail at high densities, for extinctions $A_V\gtrsim 2$-5 (e.g. \citealt{kainulainen2006,kainulainen2009,schneider2012,schneider2013} and references therein), a feature identified as the signature of gravity. Indeed, a similar feature of the PDF is found in numerical simulations of turbulence that include self-gravity (e.g. \citealt{kritsuk2010,ballesteros2011,cho2011,collins2012,federrath2013,lee2015,burkhart2016}). 

Whereas these two opposite signatures of the PDF, lognormal vs power-law, seem to be clearly identified, when and how precisely gravity starts affecting the dynamics of turbulence and thus the properties and the evolution of the PDF remains to be fully understood. Understanding the statistical properties of supersonic turbulence is at the heart of analytical theories of the star formation process, so understanding the physics governing the shape and the evolution of the density PDF of supersonic turbulence is of prime importance.
A few attempts have been made to explain the development of power law tails in the density PDF \citep{girichidis2014,Guszejnov2018,donkov2018}. These approaches, however, focus only on the gravitationally unstable parts of a cloud, using self-similar gravitational collapse models and/or geometrical arguments. While these models derive asymptotic exponents of power-law tails, they lack a complete description of the density fluctuations, both in the gravitationally stable and unstable parts of the cloud,
and treat the PDF as a static system, even though \citet{girichidis2014} follow numerically its time evolution. 
A first attempt to derive a robust theoretical framework of density fluctuations in compressible turbulence has been addressed by \citet{pan2018,Pan2019B} based on the formalism developed by \citet{pope1981,pope1985} and \citet{pope1993} for the PDF of any quantity, expressed as the conditional expectations of its time derivatives. Within the framework of this formalism, \citet{Pan2019B} used a probabilistic approach of turbulence to derive a theoretical formulation of the PDF of density fluctuations at steady state \textit{from first principles}.
In this paper, we follow a similar approach and derive an analytical theory to describe the dynamics of turbulence in dense regions of MCs and its interplay with gravity.  The approach generalizes the aforementioned ones in two ways. First, we include the impact of gravity on the cloud's dynamics. Second we consider the density PDF not as a stationary system but as a one evolving with time, implying that the conditional expectation of the flow velocity divergence is time-dependent and non zero. The theory explains the evolution of the PDF and determines the density thresholds above which gravity strongly affects and eventually dominates the dynamics of the turbulence. We provide a procedure to relate the observed {\it column density} PDFs to the underlying {\it volume density} PDFs, allowing to infer various physical parameters characterizing molecular clouds from observations. The theory and its diagnostics are confronted to numerical simulations of gravoturbulent collapsing clouds and to various available observations.

\section{Mathematical Framework} \label{sec:mathframe}
\subsection{Description of a molecular cloud}
We consider  an isolated, turbulent, self-gravitating  molecular cloud. Neglecting for now the magnetic field, the cloud's evolution is given by the standard Navier-Stokes and Poisson equations: 
\begin{eqnarray}
    \frac{\partial \rho}{\partial t}+\bm{\nabla}\cdot \left( \rho \bm{v} \right)&=&0, \label{eq::masscons} \\
    \frac{\partial \bm{v}}{\partial t}+\left(\bm{v} \cdot \bm{\nabla} \right) \bm{v}&=&- \frac{1}{\rho} \bm{\nabla}P+\bm{\mathcal{G}}+\frac{1}{\rho} \bm{\nabla} \cdot  \underline{\underline{\bm{\sigma}_\nu}}, \label{eq::momentum}\\
    \bm{\nabla}\cdot \bm{\mathcal{G}}&=&-4\pi G \rho, \label{eq::GaussNewton} 
\end{eqnarray}
where $\rho$ and $P$ denote respectively the density and pressure of the gas in the cloud, $\bm{v}$ the velocity field, $\bm{\mathcal{G}}$ the gravity field and $\underline{\underline{\bm{\sigma}_\nu}}$ the viscous stress tensor. 
We close the system of equations  by using a barotropic equation of state $P(\rho)\propto \rho^\gamma$ for the gas.

We separate the evolution of the background  from the one of local density deviations. We thus split the velocity field $\bm{v}$ between the mean velocity $\bm{V}$ and the (turbulent) velocity $\bm{u}$ (\citealt{ledoux1958}) and we introduce the logarithmic excess of local density $s$, 
\begin{eqnarray}
    \bm{V} &\equiv& \frac{1}{\overline{\rho}} \overline{\rho \bm{v}} \label{eq::meanV}, \\
    \bm{u} &\equiv& \bm{v} - \bm{V}, \label{eq::defu} \\
    \rho &\equiv& \overline{\rho}(\bm{x},t) \, \mathrm{e}^s, \label{eq:defs}
\end{eqnarray}
where $\overline{\Phi}(\bm{x},t)$ denotes the mathematical expectation, also called statistical average or mean, of any random field $\Phi$ (e.g. \citealt{pope1985,frisch1995}). We note that  $\overline{\bm{u}}\neq0$ \textit{a priori} but $\overline{\rho\bm{u}} \equiv 0$ by definition (Eqs. \ref{eq::meanV}-\ref{eq::defu}). This ensures that on average there is no transfer of mass due to turbulence and the equation of continuity (\ref{eq::masscons}) remains valid for the mean field, i.e. for $\overline{\rho}$ and $\bm{V}$.
Averaging  Eq.~(\ref{eq::masscons})--(\ref{eq::momentum}) yields an evolution equation for the mean flow written in conservative form,
\begin{eqnarray}
    \frac{\partial \left( \overline{\rho} \, \bm{V}\right)}{\partial t} +  \bm{\nabla} \cdot \left( \overline{\rho}\bm{V} \otimes \bm{V} \right)  &=& -  \bm{\nabla} \overline{P} + \overline{\rho \, \bm{\mathcal{G}}} + \nonumber \\
   &&  \bm{\nabla} \cdot \left( \overline{\underline{\underline{\bm{\sigma}_\nu}}} - \overline{\rho \bm{u} \otimes \bm{u}} \right), \label{eq::averagemomcons}
\end{eqnarray}
with all quantities replaced by their mean values, except for the appearance of the turbulent Reynolds stress tensor, $-\overline{\rho\bm{u}\otimes\bm{u}}$. The trace of this tensor corresponds to the turbulent pressure while its traceless part is related to the turbulent viscous tensor. 
We consider \textit{molecular} viscous effects to be negligible in molecular clouds and thus neglect the viscous tensor $\underline{\underline{\bm{\sigma}_\nu}}$ in the general equations.

Having obtained the averaged evolution of the cloud, we can obtain the evolution of the density deviations by substraction. 
This yields a transport equation for $s$, written in a Lagrangian form:
\begin{eqnarray}
    \frac{\mathrm{D} s}{\mathrm{D} t} = - \bm{\nabla} \cdot \bm{u} - \left(\bm{u}\cdot \bm{\nabla}\right) \rm{ln}(\overline{\rho}), \label{eq::advects} 
\end{eqnarray}
where $\frac{\mathrm{D}}{\mathrm{D} t}=\frac{\partial}{\partial t} + \left(\bm{v}\cdot \bm{\nabla}\right)$ is the Lagrangian derivative. 

\subsection{Transport equations for the probability distribution function of logarithmic density fluctuations} 
Assuming that turbulent fields are statistically homogeneous one can derive two transport equations for the probability distribution function of logarithmic density fluctuations ($s$-PDF) $f$ \citep{pope1981,pope1985,pan2018,Pan2019A,Pan2019B}:
\begin{eqnarray}
    \frac{\partial}{\partial t} f(s,t) &=& \left\lbrace 1 +\frac{\partial}{ \partial s}  \right\rbrace \left[\left< \left(\bm{\nabla}  \cdot \bm{u}\right) | s \right> f \right], \label{eq:evolvf} \\
    \frac{\partial}{\partial t}\left[ \left< \left(\bm{\nabla}  \cdot \bm{u} \right)| s \right> f \right] &=& \left\lbrace 1 +\frac{\partial}{ \partial s}  \right\rbrace \left[\left< \left(\bm{\nabla}  \cdot \bm{u}\right)^{2} | s \right> f \right] \nonumber \\ 
    &&+ f \left< \frac{\mathrm{D}\bm{\nabla}  \cdot \bm{u}}{\mathrm{D}t}| s\right>, \label{eq:dtdivu2}
\end{eqnarray}
where terms of the form $\left<\Phi | s \right> \equiv \left<\Phi | s(\bm{x},t)=s \right> $ denote the conditional expectations of the random field $\Phi$ knowing that $s(\bm{x},t)=s$, and can be computed as the average of the field $\Phi$ in all regions where $s(\bm{x},t)\in [s,s + \mathrm{d}s[$.

\subsection{Stationary solutions}\label{sec:stationary}
Eqns.~(\ref{eq:evolvf}) and (\ref{eq:dtdivu2}) give insights on the interplay between dynamical quantities and the steady state value of the density $s$PDF, $f$. \citet{pan2018,Pan2019A,Pan2019B} have
shown and tested numerically that,

1. $f$ is stationary if and only if $\left< \bm{\nabla}  \cdot \bm{u} | s \right> = 0$, $\forall s$,

2. At steady state, $f$ can be formally computed as
\begin{equation}
f(s) = \frac{C \mathrm{e}^{-s}}{\left< \left(\bm{\nabla}  \cdot \bm{u}\right)^{2} | s \right> } \mathrm{exp} \left(- \int_0^s \frac{\left< \frac{\mathrm{D}\bm{\nabla}  \cdot \bm{u}}{\mathrm{D}t}| s' \right>}{\left< \left(\bm{\nabla}  \cdot \bm{u}\right)^{2} | s' \right>} \mathrm{d}s' \right), \label{eq:pdfsteadystate}
\end{equation}
enabling us to discuss the impact of dynamical effects on the density PDF $f(s)$. 
 
\subsection{Effects of gravity on the density PDF} \label{subsec:effectsofG}

The effect of gravity on the density PDF {\it without assuming a steady state}
can  be inferred 
by recasting Eq.~(\ref{eq:dtdivu2}) as an equation for $\mathrm{ln} f$:
\begin{eqnarray}
\left<  \bm{\nabla}  \cdot \bm{u}| s \right> &&\,\frac{\partial}{\partial t} \mathrm{ln} f - \left< (\bm{\nabla}  \cdot \bm{u})^2 | s \right>   \frac{\partial}{\partial s} \mathrm{ln} f  + \frac{\partial}{\partial t} \left< \bm{\nabla}  \cdot \bm{u} | s \right>  = \left< \frac{\mathrm{D}\bm{\nabla}  \cdot \bm{u}}{\mathrm{D}t}| s \right> \nonumber \\
+ && \left\lbrace 1 +\frac{\partial}{ \partial s}  \right\rbrace \left< \left(\bm{\nabla}  \cdot \bm{u}\right)^{2} | s \right>, \label{eq:dtdivu2ln}
\end{eqnarray}
where the terms on the r.h.s. are then treated as source terms. We note that, due to Eq.~(\ref{eq:evolvf}), the term $\partial t \left< \bm{\nabla}  \cdot \bm{u} | s \right>$ on the l.h.s of Eq.~(\ref{eq:dtdivu2ln}) is in fact seen as an operator acting on $f$, in a similar way the pressure gradient is seen as a non local operator acting on the velocity field in standard studies of incompressible hydrodynamics with periodic boundary conditions (see e.g. \citealt{frisch1995}).  We then split  $\left< \frac{\mathrm{D}\bm{\nabla}  \cdot \bm{u}}{\mathrm{D}t}| s \right>$ as
\begin{equation}
    \left< \frac{\mathrm{D}\bm{\nabla}  \cdot \bm{u}}{\mathrm{D}t}| s \right> = S_{\rm turb}(s,t) + S_{\rm grav}(s,t) + S_{\rm th}(s,t) \label{eq:DtdivuCond}
\end{equation}
with 
\begin{eqnarray}
    S_{\rm grav}(s,t) &\equiv& - 4 \pi G \overline{\rho}\left(\mathrm{e}^s  - 1 \right), \label{eq:defSgrav}\\
    S_{\rm th}(s,t) &\equiv&  -\left< \bm{\nabla} \cdot \left( \frac{1}{\rho} \bm{\nabla} P \right)| s \right>, \\
    S_{\rm turb}(s,t) &\equiv& - \left< \bm{\nabla} \bm{u} : \bm{\nabla} \bm{u}| s  \right> - 2 \left< \bm{\nabla} \bm{V} : \bm{\nabla} \bm{u}| s  \right>, \label{eq:defSturb}
\end{eqnarray}
where $\bm{\nabla} \bm{x}:\bm{\nabla} \bm{y} = (\partial_i x_j) \, (\partial_j y_i)$ using  Einstein's summation convention. Eqs.~(\ref{eq:defSgrav}-\ref{eq:defSturb}) are obtained by taking the divergence of Eq.~(\ref{eq::momentum}) and subtracting its average, knowing that the turbulent fields $\rho$ and $\bm{u}$ are statistically homogeneous. Their explicit derivation is given in App. \ref{ap:DetailsSgrav}.
Then, using Eq.~(\ref{eq:dtdivu2ln}), {\it the statistics of the flow within the cloud will be dominated by gravity (i.e. will  differ from the statistics of pure gravitationless turbulence), whenever} :
\begin{equation}
    |S_{\rm grav}(s)| \gtrsim \mathrm{max}\left( |S_{turb}|, \, |S_{\rm th}|, \, \left| \left\lbrace 1 + \partial_s \right\rbrace \left< (\bm{\nabla}  \cdot \bm{u})^2 | s \right> \right|  \right). \label{eq:gravitydominantcondns}
\end{equation}
Note that if the dynamics is dominated by gravity, we expect $\left< (\bm{\nabla}  \cdot \bm{u}) | s \right> $ to be amplified in collapsing regions such that $ |S_{\rm grav}(s)| \sim \left< (\bm{\nabla}  \cdot \bm{u})^2 | s \right> $ (see \S~\ref{sec:evfreefall}). 

Physically, Eq.~(\ref{eq:gravitydominantcondns}) expresses the fact that gravity dominates whenever one of the two following conditions is fullfilled. (1) Either it overcomes thermal (pressure) or turbulent contributions to the dynamics of the cloud ($|S_{\rm grav}(s)| \gtrsim \mathrm{max}\left( |S_{turb}|, \, |S_{\rm th}|\right)$). (2) Or,  either convergent flows are produced by gravitational collapse ($|S_{\rm grav}(s)| \sim \left< (\bm{\nabla}  \cdot \bm{u})^2 | s \right>$), or divergent flows are forced to collapse, indifferently from their initial expansion ($S_{\rm grav}(s)| > \left< (\bm{\nabla}  \cdot \bm{u})^2 | s \right>$).

 As the aim of our study is to know when gravity will yield significant departures from pure (gravitationless) turbulence, we can evaluate the terms on the r.h.s. of Eq.~(\ref{eq:gravitydominantcondns}) as for standard steady-state turbulence without gravity (we denote with the subscript $\not{\! \!{\mathrm{G}}}$):
\begin{eqnarray}
    |S_{\rm grav}(s)| \gtrsim \mathrm{max}\left( |S_{turb}|, \, |S_{\rm th}|, \, \left| \left\lbrace 1 + \partial_s \right\rbrace \left< (\bm{\nabla}  \cdot \bm{u})^2 | s \right> \right|  \right)_{\not{ \rm G}}.   \quad \label{eq:gravitydominantturbless}
\end{eqnarray}
\citet{Pan2019B}  performed such an analysis 
and found that $\left< (\bm{\nabla}  \cdot \bm{u})^2 | s \right>_{\not{ \rm G}} \sim \overline{({\bm{\nabla}  \cdot \bm{u}}^2)}_{\not{ \rm G}}$, while the other terms have no straightforward functional forms.  To further simplify Eq.~(\ref{eq:gravitydominantturbless}), we start from Eq.~(\ref{eq:dtdivu2ln}) for turbulence without gravity 
\begin{eqnarray}
- \left( \left< (\bm{\nabla}  \cdot \bm{u})^2 | s \right>   \frac{\partial}{\partial s} \mathrm{ln} f \right)_{\not{ \rm G}} =  (S_{turb})_{\not{ \rm G}} + (S_{\rm th})_{\not{ \rm G}} +  \left( \left\lbrace 1 + \partial_s \right\rbrace \left< (\bm{\nabla}  \cdot \bm{u})^2 | s \right>   \right)_{\not{ \rm G}},
\end{eqnarray}
then using the triangle inequality,
\begin{eqnarray}
    \left| \left< (\bm{\nabla}  \cdot \bm{u})^2 | s \right>   \frac{\partial}{\partial s} \mathrm{ln} f \right|_{\not{ \rm G}} &\leq& |S_{turb}|_{\not{ \rm G}} + |S_{\rm th}|_{\not{ \rm G}} \nonumber \\
    && +  \left| \left( \left\lbrace 1 + \partial_s \right\rbrace\left< (\bm{\nabla}  \cdot \bm{u})^2 | s \right>   \right)_{\not{ \rm G}} \right| ,
\end{eqnarray}
yields for the condition given by Eq.~(\ref{eq:gravitydominantturbless}):
\begin{equation}
    |S_{\rm grav}(s)| \gtrsim \left| \left< (\bm{\nabla}  \cdot \bm{u})^2 | s \right>   \frac{\partial}{\partial s} \mathrm{ln} f \right|_{\not{ \rm G}}. \label{eq:criteriongravity}
\end{equation}
Making the standard approximation that $f_{\not{ \rm G}}$ is a lognormal form of variance $\sigma_s$ yields the simplified condition:
\begin{equation}
   \boxed{ |S_{\rm grav}(s)| \gtrsim \overline{(\bm{\nabla}  \cdot \bm{u})^2}_{\not{ \rm G}} \times \left| \frac{s + \frac{1}{2}\sigma_s^2}{\sigma_s^2}\right| }\, , \label{eq:gravitythreshold}
\end{equation}
where  $\sigma_s$ is given in terms of the rms Mach number $\mathcal{M}$ and forcing parameter $b$ as (e.g. \citealt{federrath2008}): 
\begin{equation}
 \sigma_s^2 = \mathrm{ln}(1 + (b \mathcal{M})^2). \label{eq:mach-variance}
\end{equation}
Then, approximating $\overline{(\bm{\nabla}  \cdot \bm{u})^2}_{\not{ \rm G}}$  within an order of magnitude estimate as:  
\begin{equation}
    \overline{(\bm{\nabla}  \cdot \bm{u})^2}_{\not{ \rm G}} = \frac{1}{\overline{\rho}^2} \frac{\overline{(\Delta \rho)^2}}{\tau_{\rm turb}^2},
\end{equation}
where $\overline{(\Delta \rho)^2} = \overline{(\rho - \overline{\rho})^2} \simeq (b \mathcal{M} \overline{\rho})^2$ and $\tau_{\rm turb}$ is a typical turbulent timescale, of the order of the crossing time $\tau_{\rm c} = L_{\rm c}/(2 \sigma_v)$, with $\sigma_v $ the 3D velocity dispersion and  $L_{\rm c}$ the diameter of the cloud, Eq.~(\ref{eq:gravitythreshold}) reduces to
\begin{empheq}[box=\fbox]{align}
     |\mathrm{e}^s - 1| &\gtrsim& (b \mathcal{M})^2 \times \left(\frac{\tau_{\rm G,0}}{\tau_{\rm turb}}\right)^{2} \times \left| \frac{s + \frac{1}{2}\sigma_s^2}{\sigma_s^2} \right| \nonumber \\
    &\gtrsim& (b \mathcal{M})^2 \times \alpha_{\rm vir}(t) \times \left| \frac{s + \frac{1}{2}\sigma_s^2}{\sigma_s^2}\right| , \label{eq:gravitythresholdvir}
\end{empheq}
where $\tau_{\rm G,0} = 1/\sqrt{4 \pi G \overline{\rho}}$ and $\alpha_{\rm vir}(t)= 5 \sigma_v ^2 /( \pi G L_{\rm c}^2 \overline{\rho} (t))$ is the virial parameter, equal to $=2E_{\rm kin}/E_{\rm grav}$ for a homogeneous spherical cloud. 

\noindent This equation introduces a {\it new characteristic timescale},  $\tau_{\rm G,0} \equiv 1/\sqrt{4 \pi G \overline{\rho}}$.
    This timescale characterizes the impact of gravity  upon turbulence in the PDF evolution of the cloud, as formalized by Eqns.~(\ref{eq:DtdivuCond}) and (\ref{eq:gravitythresholdvir}). It it is roughly half the mean free-fall time of the cloud, $ \tff \equiv   \sqrt{\frac{3 \pi}{32 G \overline{\rho}}}$.
{\it Equation~(\ref{eq:gravitythresholdvir}) then allows  a determination, within a factor of a few,  of the density above which gravity is expected to change significantly the statistics of turbulence}. 

Furthermore, following \citet{Pan2019B} and using Eq.~(\ref{eq:pdfsteadystate}), we see that because $S_{\rm grav}(s) < 0$ when $s > 0$ and $S_{\rm grav}(s) > 0$ when $s<0$, respectively, gravity  tends to broaden the PDF both at small and large densities, {\it resulting in a larger variance compared with the case with no gravity}. This can be understood by considering that gravity acts as an extra compressive forcing.  This is equivalent to increasing 
($b \mathcal{M}$) 
in compressible turbulent simulations (Eq. (\ref{eq:mach-variance})).

Therefore, according to the present analysis, we expect to have typically 2 regions (in terms of density) with different contributions governing the statistics of turbulence: 
    
\indent - a {\it first region}, corresponding to  $s < s_{ \rm G}$, where $s_{ \rm G}$ is  given by Eq.~(\ref{eq:gravitythresholdvir}), where the statistics is similar to the one of gravitationless turbulence but with a (more or less)
increased variance due to gravity. The $s$-PDF in this region is lognormal-like,
    
    \indent -  a {\it second region}, corresponding to densities  $s > s_{ \rm G}$, where gravity has a dominant impact on the statistics of turbulence, and the PDF will depart from (gaussian) lognormal statistics.
     
 The threshold density, $s_{\rm G}$, between the two regions {\it evolves with time on the same timescale $\tmean$ as the global, average} properties of the cloud ($\overline{\rho}(t)$, $\alpha_{\rm vir}(t)$, $\overline{(\bm{\nabla}  \cdot \bm{u})^2}(t)$)\footnote{\label{foot:timescale}The timescale $\tmean$ of variation of $\overline{\rho}$ is not necessary equal to $\tff$. If there is enough turbulent support for example, it can be larger. It depends on what drives the \textit{global} evolution of the cloud.}, according to Eq.~(\ref{eq:gravitythresholdvir}). However, as will be shown in \S \ref{sec:evolutionPDF},  at densities $s > s_{\rm G}$, the PDF will start departing from a lognormal form and develop a power law on {\it shorter timescales}, of the order of a typical \textit{local} free-fall time, $\tau_{\rm ff}(s) < \tmean$. 



\section{Evolution of the density PDF in star forming clouds} \label{sec:evolutionPDF}

Observations of column-density PDFs in MCs  show that regions where star formation has not occured yet exhibit lognormal PDFs whereas regions with numerous prestellar cores exhibit power-law tails at high column densities \citep{kainulainen2009,schneider2013}. Similarly, numerical simulations of star formation in turbulent clouds  show that density PDFs develop power-law tails as the simulations evolve  \citep{klessen2000,federrath2013}. This suggests that the density PDF in star forming clouds is not stationary but evolves with time, implying  $\left< \bm{\nabla}  \cdot \bm{u} | s \right> \neq 0$ (see \S\ref{sec:stationary}). 

\subsection{Mathematical derivation: equivalence of the velocity divergence and $s$-PDF power-law tail exponents.}
Finding solutions of Eq.~(\ref{eq:evolvf}) for any function $\left< \bm{\nabla}  \cdot \bm{u} | s, t \right>$ is not straightforward. Assuming, as a simplification, separability of the time and density variables
    $\left< \bm{\nabla}  \cdot \bm{u} | s, t \right> = h(t) \times g(s)$
yields, from the method of characteristics, the solution 
\begin{equation}
    f(s,t) = \Phi \left( \int h \mathrm{d}t' + \int \frac{1}{g} \mathrm{d}s' \right) \frac{\mathrm{e}^{-s}}{g(s)},
\end{equation}
with $\Phi$  any differentiable function.  

We prove now that a non-stationary $s$-PDF develops a power-law tail of exponent $\alpha_s=a+1$, with $a>0$, if and only if the conditional expectation of the velocity divergence $\left< \bm{\nabla}  \cdot \bm{u} | s \right>$ scales at large $s \gg 1$ as $\left< \bm{\nabla}  \cdot \bm{u} | s, t \right> \approx h(t)  \times \mathrm{e}^{a s}$. Indeed, if for $s \geq s_c$,  for some $s_c$, one has
\begin{equation}
    \left< \bm{\nabla}  \cdot \bm{u} | s, t \right> = h(t) \times \mathrm{e}^{a s},
    \label{eq:separ}
\end{equation}
with $a>0$, then
\begin{equation}
    f(s,t) = \Phi \left( a  \int h \mathrm{d}t' - \mathrm{e}^{-a s}  \right) \mathrm{e}^{-(1+a)s}, \label{eq:solfexp}
\end{equation}
for $s>s_c$, from Eq.~(\ref{eq:separ}). Hence, as  $a>0$, the PDF is expected to develop a power-law tail  with an exponent $-(1+a)$
at a given time $t$ at $s>s_c$ sufficiently large  such that  $\Phi \left( a  \int h \mathrm{d}t' - \mathrm{e}^{-a s}  \right) \approx \Phi \left( a  \int h \mathrm{d}t'  \right)$. The proof of the reciprocal of this result (Eqs \ref{eq:separ}-\ref{eq:solfexp}) is given in Appendix \ref{app:reciprocal}. Thus, observed power-law tails, $f(s,t) \propto e^{-\alpha_s s}$, with exponents $\alpha_s = 3/2$ and $2$ correspond to an underlying expectation $\left< \bm{\nabla}  \cdot \bm{u} | s, t \right>  \propto e^{s/2}$ and $\left< \bm{\nabla}  \cdot \bm{u} | s, t \right>  \propto e^{s}$, respectively.

\subsection{Physical interpretation: Transitory regime and short time evolution} \label{sec:intemer}

At any time in a cloud, we can compute the threshold value $s_{\rm G}$ above which gravity starts altering significantly the statistics of fully developed turbulence. For diffuse, hot and/or turbulent clouds ($\alpha_{\rm vir} \gg 1$), however, this value can be so large that the probability $\mathcal{P}(s>s_{ \rm G})$ of finding regions $s>s_{ \rm G}$, becomes very small. In such cases, one can completely neglect the effect of gravity. To be more quantitative, let us assume that gravity can be neglected if $\mathcal{P}(s>s_{ \rm G}) \leq 10^{-9}$. Assuming a lognormal PDF, this yields $s_{\rm G}  \geq 6 \sigma_s - 0.5 \, \sigma_s^2$ (where $\sigma_s$ is the variance in Eq.~(\ref{eq:mach-variance})). In hot and turbulent clouds, where $T \sim 8000$ K, $(b\mathcal{M}) \sim 1$ (e.g. Draco, \citealt{miville2017}) this yields $\alpha_{\rm vir} \gtrsim 5.5$ from Eq. \ref{eq:gravitythresholdvir}. As the cloud cools down and contracts,  $\alpha_{\rm vir}$  decreases, resulting in a small enough value of $s_{\rm G}$ to observe significant departures from a lognormal PDF. We can thus define a time $t_0$ in the lifetime of the cloud as the time at which  the volume fraction of the cloud corresponding to (dense) regions with $s>s_{\rm G}$, where the gas PDF starts departing from the statistics of pure turbulence, $ \left< \bm{\nabla}  \cdot \bm{u} | s\right> \simeq 0$, under the growing influence of gravity, becomes {\it noticeable}, i.e. statistically significative.  This fixes the ‘‘zero of time" in star forming cloud lifetimes, whatever the (undefinable) time since which they have been formed.  The time $t_0$ thus corresponds to the time at which some dense regions start to collapse and depart from the \textit{global} evolution (contraction or expansion) of the cloud, which is described by the time variation of $\overline{\rho}(t)$. This time $t_0$ then enables us to determine a physically motivated value to fix the indefinite integral in Eq. \ref{eq:solfexp}, as the one being equal to $0$ at $t_0$.

For regions with $s>s_{\rm G}$, we expect from Eqs.~(\ref{eq:dtdivu2ln}-\ref{eq:DtdivuCond})
 at short times $t= t_0 + \tcoll$ after $t_0$, i.e. in the linear regime, to have $\left< \bm{\nabla}  \cdot \bm{u} | s \right> \simeq - 4 \pi  G \overline{\rho}\mathrm{e}^s (t-t_0) =- 4 \pi  G \overline{\rho}\mathrm{e}^s  \tcoll$ (i.e. $a=1$ in  Eq.~(\ref{eq:solfexp})), yielding  for the PDF $f$:

\begin{equation}
    f(s,t) \simeq \tilde{\Phi}\left(\tau_{\rm G,0}^{-2} \,  \frac{\tcoll^2}{2} + \mathrm{e}^{-s} \right) \mathrm{e}^{-2s}. \label{eq:transitoryfreefall}
\end{equation}

Therefore, for densities $s>s_{\rm G}$, we  expect to see the onset of a \textit{first power law tail} in the $s$-PDF, $f(s,t)\propto e^{-\alpha_s s}$, with a steep exponent $\alpha_s \simeq 2$ in a typical timescale $ \tau_{\rm G}(s)= \tau_{\rm G,0} \,\mathrm{e}^{-s/2} $. As seen from Eq.~(\ref{eq:transitoryfreefall}),  the onset of this first power-law tail occurs, for a given time $t$, at a density ${s_t}=\rho_t (t_{\rm coll})/\overline{\rho}\simeq (\tau_{\rm G,0}/t_{\rm coll})^2 
\simeq 0.25 (\tff/t_{\rm coll})^{2}$, as found in numerical calculations \citep{girichidis2014}. At later time (a few $\tau_{\rm ff}(s)$, see \S \ref{sec:evfreefall}) for a given density or at higher densities for a given time, a \textit{second power-law} develops with $\alpha_s=3/2$, signature of regions in free-fall collapse, as seen in \S\ref{sec:evfreefall}. 


\subsection{ Asymptotic case: evolution in regions of  ``free-fall" collapse} \label{sec:evfreefall}

The densest regions  in star forming clouds are expected to collapse under their own gravity on a timescale of the order of the {\it local free-fall time} $\tau_{\rm ff}(\rho)\propto (G \rho)^{-1/2}$. For these regions  we thus expect a scaling:
\begin{equation}
   - \left< \tau_{\rm ff}^{-1}(\rho) | s \right> \propto \left< \bm{\nabla}  \cdot \bm{u} | s \right> = - c \sqrt{4 \pi G\overline{\rho}} \, \mathrm{e}^{s/2}, \label{eq:scalingdivufreefall}
\end{equation}
where $c$ is a constant of proportionality of order unity. This yields, from Eq.~(\ref{eq:solfexp}): 
\begin{equation}
  f(s,t) = \Phi \left(\frac{c}{2}  \sqrt{4 \pi G} \int_{t_0}^t \sqrt{\overline{\rho}(t')} \, \mathrm{d}t'  +  \mathrm{e}^{- s/2}  \right) \mathrm{e}^{-\frac{3}{2}s},
\end{equation}
where $t = t_0 + \tcoll$.
Then, if the time after which a dense region of the cloud started to collapse, $\tcoll$, is short compared to the characteristic time of variation of $\overline{\rho}$ 
, $\tcoll\ll \tmean$, meaning that the global properties of the cloud did not have time to evolve significantly, we can write
\begin{equation} 
    f(s,t) \simeq \Phi \left( \frac{c}{2}  \sqrt{4 \pi G \, \overline{\rho}(t)}\, \tcoll   +  \mathrm{e}^{- s/2}  \right) \mathrm{e}^{-\frac{3}{2}s}.\label{eq:freefall} 
\end{equation}
Therefore, the PDF develops a power-law tail with {\it a specific exponent} $-3/2$ for $s\ge s_{ \rm G} $ within a typical time $t(s) \equiv 2\, c^{-1} \, \tau_{\rm G,0} \,\mathrm{e}^{-s/2} \simeq c^{-1} \tff \,\mathrm{e}^{-s/2}  \simeq c^{-1} \tau_{\rm ff}(s) $.  

This  analysis thus shows  that the onset of power-law tails, $f(s)\propto e^{-\alpha_s s}$, in the PDF reflects the growing impact of gravity on the turbulent flow, with a first power-law exponent $\alpha_s \lesssim 2$, reaching the asymptotic value $\alpha_s=3/2$ in  free-fall collapsing  regions.

\section{From volume to column densities } \label{sec:columndens}

Observations of dense MCs trace  the density integrated along the line of sight, and thus reveal the PDF of the column density $\Sigma$ or its logarithmic deviations $\eta= \mathrm{ln}(\Sigma/\overline{\Sigma})$. Much efforts have been made to link the observed $\eta$-PDF to properties of the underlying $s$-PDF \citep{Vazquez_Semadeni_2001,brunt2010,Burkhart2012,federrath2013}. In the present study, we will use the relation of \citet{Burkhart2012} to illustrate our findings.
%
%
Furthermore, we will adopt the relation given by \citet{federrath2013} to link the exponents $\alpha_\eta$ and $\alpha_s$ of the $\eta$-PDF and $s$-PDF, respectively:
\begin{equation}
    \alpha_\eta = - \frac{2}{1-\frac{3}{\alpha_s}}. \label{eq:relationalpha}
\end{equation}
Hence, for regions in free-fall collapse we expect a power-law tail in the $\eta$-PDF with an asymptotic exponent $\alpha_\eta = 2$ ($\alpha_s = 3/2$), with a transition domain with $\alpha_\eta \geq 4 $ ($\alpha_s \geq 2$) (see \S \ref{sec:evolutionPDF}).  

In order to make comparison between our theory and numerical simulations or observations for the $\eta$-PDF, we have derived  a way to relate the volume density at which the $s$-PDF, $f(s)$, develops power-laws to the column density at which the $\eta$-PDF, $p(\eta)$, develops a similar behaviour. We call  $s_{\rm crit}$ and  $\eta_{\rm crit}$ the  critical value corresponding to the beginning of a power-law tail in the two respective PDFs.  Assuming ergodicity and statistical isotropy we obtain (see App. \ref{app:columndens} for details)
\begin{equation}
    \int_{\eta_{\rm crit}}^\infty p(\eta) \mathrm{d} \eta  \simeq \left( \int_{s_{\rm crit}}^\infty f(s) \mathrm{d}s \right)^{2/3}. \label{eq:linkedcritical}
\end{equation}
In case there are 2 power-law tails, starting at $s_{1}$ and $s_2$, Eq.~(\ref{eq:linkedcritical}) remains a good approximation as long as $s_2 - s_1 \gtrsim 1$, so that the upper bound in the integrals has not much importance. This procedure is tested against numerical simulations in \S \ref{sec:numerical} and confronted to observations in \S \ref{sec:obs}. More details on how one obtains Eq.~(\ref{eq:linkedcritical}) are given in App. \ref{app:columndens}.

\section{Comparison with numerical simulations} \label{sec:numerical}

\subsection{Numerical set up}

To understand how gravity affects the $s$ and $\eta$ PDFs in star forming clouds and compare with our theoretical formulation, we use the numerical simulations of isothermal self-gravitating turbulence on 3D periodic grids presented in \citet{federrath2012,federrath2013}, kindly provided by the authors.
These simulations model isothermal self-gravitating magneto-hydrodynamic turbulence on 3D periodic grids with resolution $N_{\rm res}^3=128^3$ to $1024^3$. Here, we will only consider simulations with no magnetic field. In the simulations, turbulence is driven with solenoidal or compressive forcing or with a mixture of both. Sink particles are used (see \citet{federrath2012} or our App. \ref{app:numericalmodels} for  details).

\begin{figure*}[!ht]
    \centering
    \includegraphics[width=\textwidth]{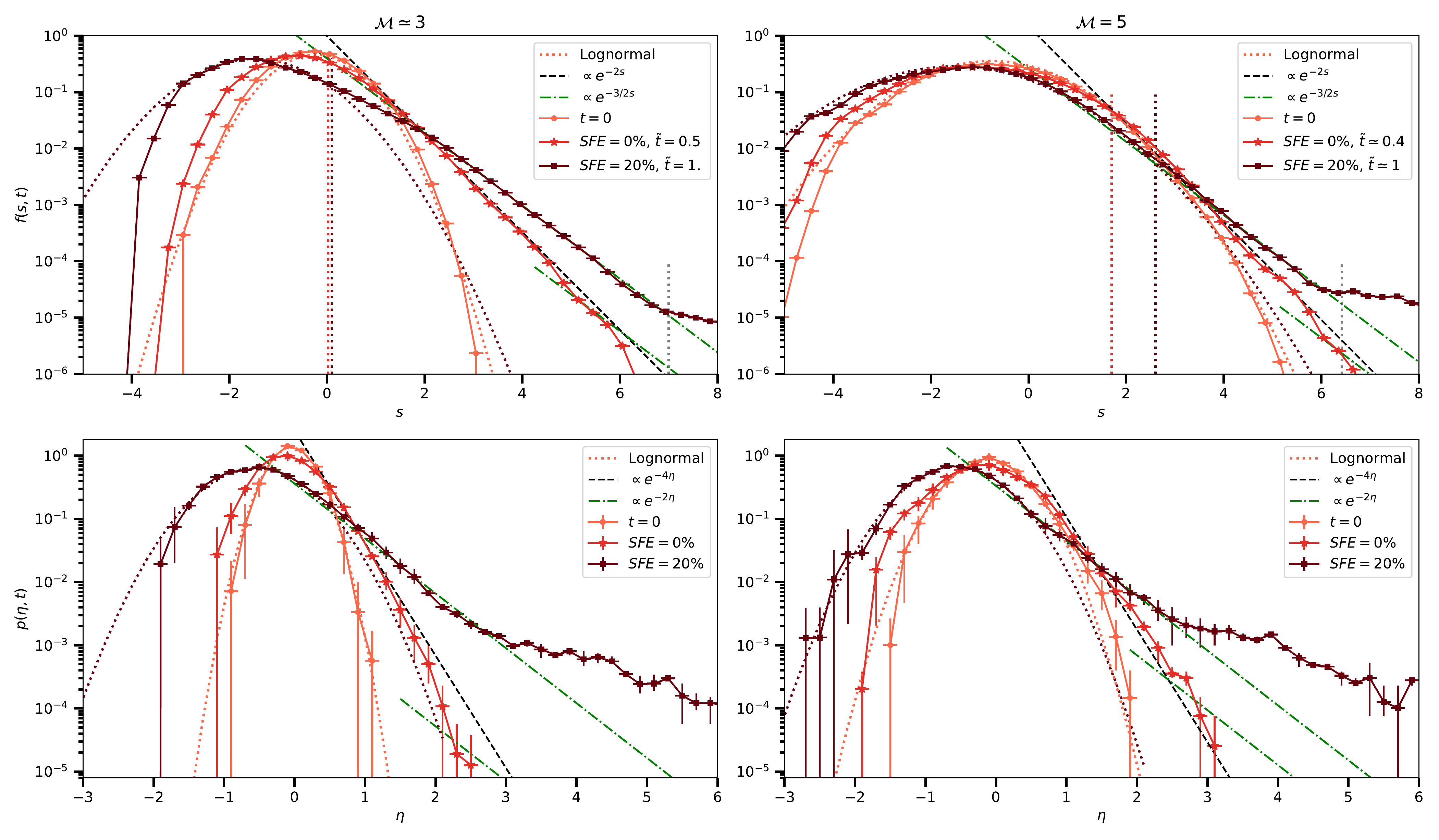}
    \caption{Evolution of the $s$ (top row) and $\eta$-PDFs (bottom row) for solenoidal simulations with $\mathcal{M} \simeq 3$, $N_{\rm res}=512$ (left) and $\mathcal{M}=5$, $N_{\rm res}=256$ (right) at $t=0$ (orange circles), SFE$=0\%$ (red stars) and SFE$=20\%$ (dark red squares). Horizontal error-bars represent bin spacing and vertical error-bars indicate the uncertainty in the $\eta$-PDFs corresponding to 3 different projection directions of the simulation box. Lognormal fits of the low density parts of PDFs at $t=0$ and SFE$=20\%$ are shown in dotted lines.The vertical dotted red lines corresponds to the value of $s_{\rm G}$ calculated from Eq. (\ref{eq:simulationsg}) with values of $(b\mathcal{M})$ and $\sigma_s$ calculated at time $\tilde{t} = t/\tff$. For $s>s_{\rm G}$ , the $s$-PDFs and $\eta$-PDFs first develop power-law tails with exponents $\alpha_\eta=2$, $\alpha_\eta=4$  (black dashed lines) and then $\alpha_s = 3/2$, $\alpha_\eta=2$ (green dot-dashed lines) at higher density.  The vertical dotted grey lines at $s>6$ correspond to $s_{\rm max}$ from Eq. (\ref{eq:smaxvir}).}
     \label{fig:numericalPDFs}   
\end{figure*}

%
%
After Eq.~(\ref{eq:gravitythresholdvir}), we expect gravity to have a dominant contribution at densities $s>s_ {\rm G}$, which yields here:
\begin{eqnarray}
     |\mathrm{e}^{s_{\rm G}} - 1| &\equiv& (b \mathcal{M})^2 \times  \alpha_{\rm vir,0} \times \left|\frac{s_{\rm G} + \frac{1}{2}\sigma_s^2}{\sigma_s^2}\right|, \label{eq:simulationsg}
\end{eqnarray}
where $\alpha_{\rm vir,0} = 5 \sigma_v ^2 /(6 G L_{\rm b}^2 \rho_{0})$ is the virial parameter suited for a box of size $L_{\rm b}$ and 3D-velocity dispersion $\sigma_v$, as in \citet{federrath2012,federrath2013}\footnote{Note that this differs from the $\alpha_{\rm vir,0}$ defined in \S \ref{subsec:effectsofG} by a factor $\pi/6 \simeq 1/2$, if the cloud size $L_{\rm c}$ is taken to be the box size $L_{\rm b}$}.  On the other hand, the maximum density $ \rho_{\rm max}$ above which the simulations do not properly resolve the collapse and describe the statistics of the cloud reads \citep{truelove1997}:
\begin{equation}
    \rho_{\rm max} = \rho_0 \, \mathrm{e}^{s_{\rm max}}=\frac{\pi c_{\rm s}^2}{16 G \Delta x_{\rm min}^2}, \label{eq:Truelovecs}
\end{equation}
with $\Delta x_{\rm min}$ the size of the most resolved cell. This condition can be rewritten
\begin{eqnarray}
    s_{\rm max} =   \mathrm{ln}\left(\alpha_{\rm vir,0} \right) + 2 \, \mathrm{ln}\left( \frac{N_{res}}{\mathcal{M}}\right) + \mathrm{ln}\left(\frac{6 \pi}{80}\right). \label{eq:smaxvir}
\end{eqnarray}

\noindent For $s>s_{\rm max}$, the lack of resolution will yield the development of shallow power-laws corresponding to the spurious fragmentation of these regions (see Fig.~(\ref{fig:numericalPDFs}) below and \citet{federrath2013}).
\subsection{Evolution of the PDFs} \label{subsec:pdfevolution}

\begin{figure}[!ht]
    \centering
    \includegraphics[width=\columnwidth]{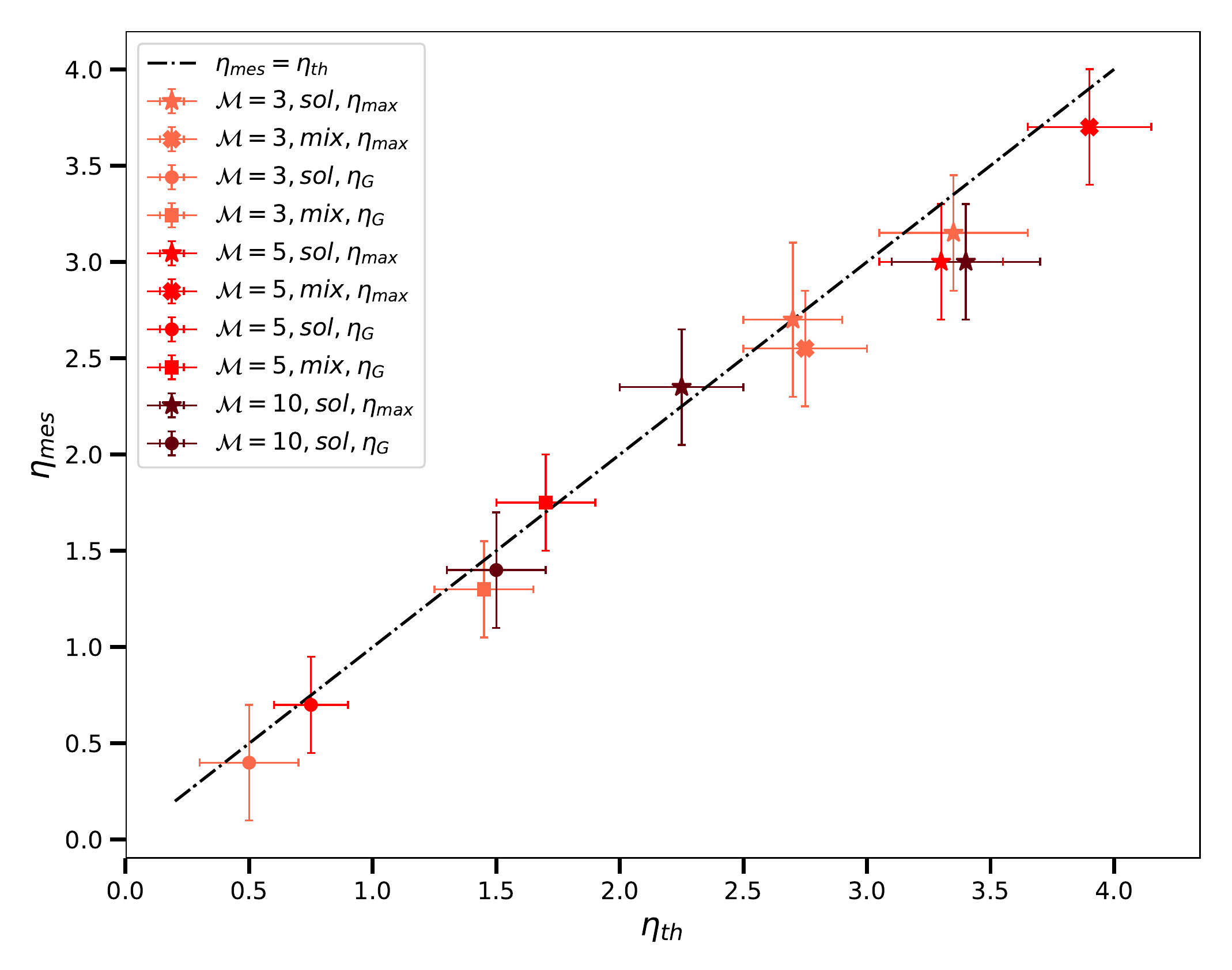}
    \caption{$\eta_{\rm th}$, calculated from  Eq.~(\ref{eq:linkedcritical}) and the $s$-PDF, vs.  $\eta_{\rm mes}$, directly measured on the $\eta$-PDFs, for runs with $\mathcal{M} = 3$, $5$, $10$  (light to dark red). Runs with $\mathcal{M} \sim 3$ and $\mathcal{M}=10$ have two values of $\eta_{\rm max}$, for each resolution $N_{\rm res} =256$ and  $N_{\rm res} =512$. We note the excellent agreement between  $\eta_{\rm mes}$ and $\eta_{\rm th}$.}
    \label{fig:sgetag}   
\end{figure}
%

Figure~(\ref{fig:numericalPDFs}) compares our analytical calculations of the $s$ and $\eta$-PDFs with the solenoidal simulations for $\mathcal{M} \simeq 3$, $N_{\rm res}=512$ and $\mathcal{M}=5$, $N_{\rm res}=256$ at initial time $t=0$ (orange), and at star formation efficiencies SFE$=$0\% (red) and SFE$=$20\% (dark red).
A major result of  \S~\ref{subsec:effectsofG} (Eqns.~(\ref{eq:gravitythresholdvir}) and
(\ref{eq:simulationsg})) is the determination of a density threshold, $s_{\rm G}$ (resp. $\eta_{\rm G}$),  above which the $s$-PDF (resp. $\eta$-PDF) is expected to develop power-law tails.  Similarly, spurious shallow power-laws will develop above $s_ {\rm max}$ (resp. $\eta_ {\rm max}$). In both cases $\eta_{\rm G}$ and $\eta_{\rm max}$ can be obtained from the determination of the corresponding values on the $s$-PDF with Eq.~(\ref{eq:linkedcritical}).  We note the excellent agreement between the theoretical and numerical curves over the whole range of densities. Notably, the theoretical determinations of $s_{\rm G}$ from Eq.~(\ref{eq:simulationsg}), threshold of the gravity impacted domain, agree very well with the onset of a power law in the simulations. It is worth stressing that for the $\mathcal{M} \sim 3$ simulation, $s_{\rm G} \sim 0.1$ and $\vir \ll 1$, and thus we do not expect the power law tail with exponent $\alpha_s = 3/2$ to develop up to $s_{\rm G}$ in a time $\tilde{t} \simeq 1$ (since this requires typically a few $\tff$). However departures from a lognormal behavior are indeed seen to start at about $s \sim 0.1$.

\noindent Fig.~(\ref{fig:sgetag}) compares $\eta_{\rm mes} \equiv \eta_{\rm G_{\rm mes}},\, \eta_{\rm max_{\rm mes}}$, directly measured on various simulations, to $\eta_{\rm th} \equiv \eta_{\rm G_{\rm th}},\, \eta_{\rm max_{\rm th}}$ derived from Eq.~(\ref{eq:linkedcritical}) and the value of $s_{\rm G}$ and $s_{\rm max}$. As seen in the figure, the agreement between the theoretical value $\eta_{\rm th}$ and the measured  one $\eta_{\rm mes}$ is remarkable.

\section{Comparison with Observations} \label{sec:obs}

In this section, we confront our theory to observations of column density PDFs in various MCs.
We use a simple model with one or two power-law tails, characterized by 1 or 2 transition densities, $s_1$ and $s_2$, between lognormal and power laws, as described in App.~(\ref{app:modelPLT}).
From the determination of the variance $\sigma_{\rm s, \eta}$ in the lognormal parts of the PDF, we get an estimate of the product  $(b \mathcal{M})$ (Eq.~(\ref{eq:mach-variance})), while from the determinations of $s_1$ and $s_2$ we get an estimate of $\alpha_{\rm vir} = 5 \sigma_v ^2 /( \pi G L_{\rm c}^2 \overline{\rho}) $, and of the time since the first regions started to collapse, in unit of mean free-fall time $\tilde{t}_{\rm coll}={t}_{\rm coll}/\tff$ (Eqs.~(\ref{eq:gravitythresholdvir}), (\ref{eq:freefall}), (\ref{eq:transitoryfreefall})). The values are given in Table (\ref{tab:obs}).

\begin{figure*}[!ht]
    \centering
    \includegraphics[width=\textwidth]{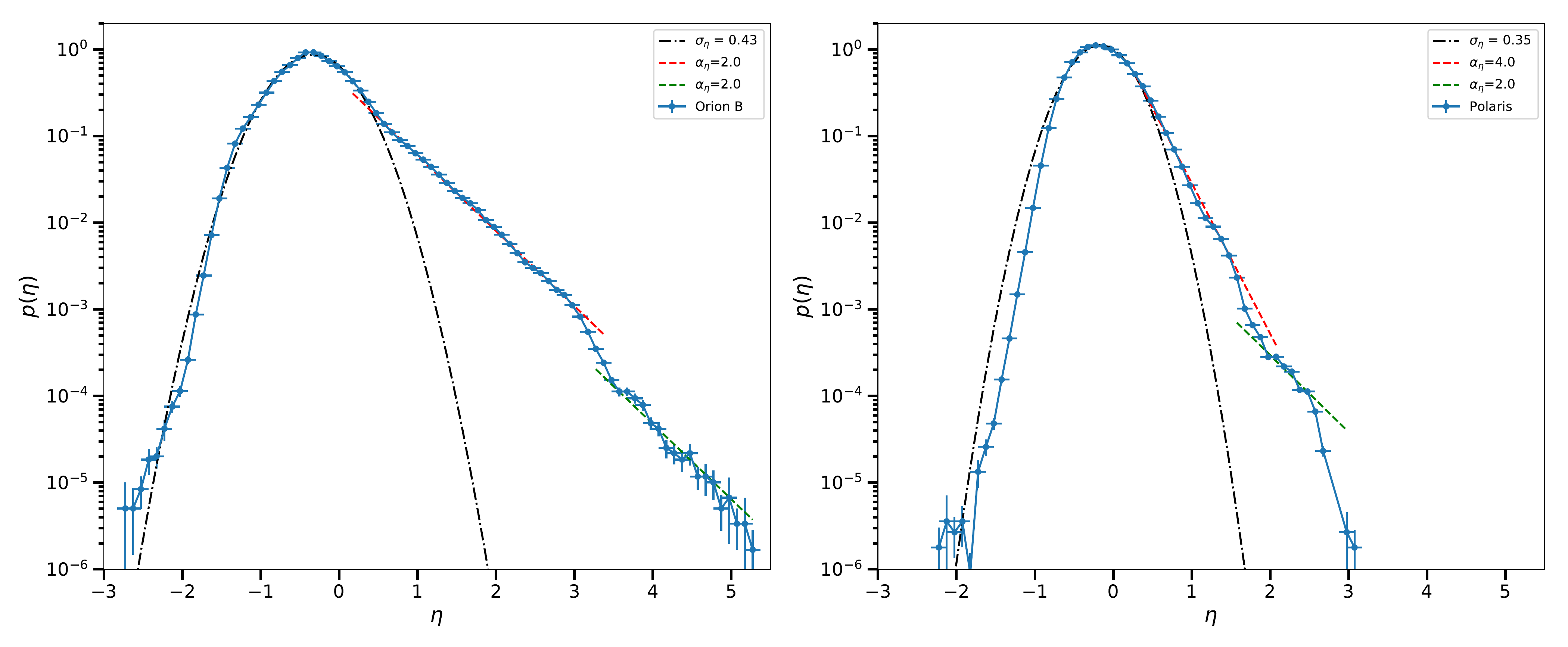}
    \caption{Left: Observed $\eta$-PDF of the cloud Orion B. Dash-dotted black line: lognormal fit of the low $\eta$ part of the PDF. Dashed red and green lines: power-law with exponent $\alpha_\eta = 2$, corresponding to an underlying $s$-PDF with a power-law exponent $\alpha_s=3/2$, signature of collapsed regions. Right: Observed $\eta$-PDF of the cloud Polaris. Dash-dotted black line: lognormal fit of the low $\eta$ part of the PDF. Dashed red and green lines: power-laws respectively with exponents $\alpha_\eta =4$ and 2, corresponding to an underlying $s$-PDF transiting from a lognormal to power laws with exponents $\alpha_s=2$ and 3/2, respectively.}
    \label{fig:obs}   
\end{figure*}


 We apply our method to two different clouds: Orion B \citep{schneider2013,orkisz2017} and Polaris \citep{andre2010,miville2010,schneider2013}.  The data were kindly provided by Nicola Schneider. For Orion B, the average column density is $\overline{N(H_2)} = 2.06 \times 10^{21} \mathrm{cm}^{-2}$ and the cloud's total mass and area above an extinction $A_v \geq 1$ are $M_{c, A_v \geq1}=29.69 \times 10^3 M_\odot$ and $A_{c, A_v \geq1} = 651$ pc$^2$. For Polaris this yields $( \overline{N(H_2)}, M_{c, A_v \geq1}, A_{c, A_v \geq1}) = (1.73 \,10^{21} \mathrm{cm}^{-2}, 1.21 \times 10^3 M_\odot, 3.9\, \mathrm{pc}^2)$.

The first one, Orion B, contains numerous pre-stellar cores. Its $\eta$-PDF displays a lognormal part at low column densities and a power-law tail at high densities with exponent $\alpha_\eta \simeq 2$, corresponding to an underlying $s$-PDF with exponent $\alpha_s=3/2$, signature of collapsed regions, as seen in Fig.~(\ref{fig:obs}) (left).  The power-law tail develops for $s>s_1 = 1.73^{+0.25}_{-0.23}$. We can thus estimate that in this cloud, (statistically significant, see \S3) 
collapse of the densest regions has occured since $\tilde{t}_{\rm coll}\approx (2-5)\times \mathrm{e}^{-s_{1}/2} \gtrsim 1$ (see \S \ref{sec:evfreefall}). Note here that $\tff$ corresponds  to the {\it region  under study} in the cloud, not to the global cloud itself. Estimation of $(b \mathcal{M})$  from the determination of $\sigma_s$ combined with estimation of $s_1$ yield $\alpha_{\rm vir} \sim 1$.  The estimated $(b \mathcal{M})$ is compatible with mean Mach-numbers $\mathcal{M} = 5.4^{+0.8}_{-0.8}$ for $b=1/3$ and $\mathcal{M} = 3.6^{+0.5}_{-0.5}$  for $b=1/2$, in agreement with \citet{orkisz2017}.
 \begin{table*}[!t]
    \centering
    \caption{Properties of the clouds}
    \begin{tabular}{c c c c c c c c c c c}
    \hline
    \hline
     Name & Func. form &$\sigma_{s}$ & $(b\mathcal{M})$ & $\mu $ & $\alpha_{1}$ & $\alpha_2$ & $s_{1}$   & $s_{2}$ & $\alpha_{\rm vir}$ & $\tilde{t}_{\rm coll}$  \\
     (1)  &  (2)         &  (3)     & (4)              &    (5)        & (6)  & (7) & (8) & (9) & (10) & (11)\\
     \hline
     Orion B & Ln+1Pl  &  $1.2^{+0.09}_{-0.1}$ & $1.8^{+0.26}_{-0.26}$ & $-0.92^{+0.13}_{-0.11}$ & $2$  & $-$  & $1.73^{+0.25}_{-0.23}$ & $-$ & $0.88^{+0.26}_{-0.23}$ &$\gtrsim 1$ \\
      Polaris & Ln+2Pl  &  $0.98^{+0.07}_{-0.08}$   & $1.27^{+0.16}_{-0.15}$ & $-0.55^{+0.08}_{-0.07}$ & $4$ & $2$  & $1.68^{+0.38}_{-0.34}$ & $6.3^{+0.1}_{-0.15}$ & $\lesssim 1.2$ & $0.2 \pm 0.1$\\
      \hline
      \hline
    \end{tabular}
    \tablecomments{Columns: (1) cloud's name; (2) functional form: Ln+1Pl, Ln+2Pl (Lognormal and 1 power law, Lognormal and 2 power laws); (3) standard deviation of the lognormal part $\sigma_s$; (4) $(b \mathcal{M})$ associated to $\sigma_s$; (5) most probable $s$-value  $\mu$; (6) exponent of the first power-law $\alpha_1$; (7) exponent of the second power-law $\alpha_2$; (8) transition between the Lognormal part and the first power-law $s_1$; (9) transition between the two power-laws $s_2$; (10) virial parameter $\alpha_{\rm vir}$ associated to $s_1$; (11) time since the first region started to collapse in units of mean free-fall time, $\tilde{t}_{\rm coll}$. }
    \label{tab:obs}
\end{table*}

The second cloud, Polaris, where detectable star formation does not seem to have occurred yet, exhibits an extended power-law tail with a steep exponent,  $\alpha_\eta \simeq 4$, corresponding to a $s$-PDF power-law tail of exponent $\alpha_s=2$ for $s>s_1$, before reaching the asymptotic values $\alpha_\eta \simeq 2$, i.e. $\alpha_s=3/2$ at high density, $s>s_2$, as seen in Fig.~(\ref{fig:obs}) (right). Carrying out the same analysis as for Orion B, we get $(s_1,s_2) = (1.68^{+0.38}_{-0.34},6.3^{+0.1}_{-0.15})$.  The value of $s_1$ yields here $\tilde{t}_{\rm coll} \approx 0.5 \, \mathrm{e}^{-s_{1}/2} =0.22^{+0.03}_{-0.04}$ for this cloud. The determination of the density $s_2$, which corresponds to collapsing regions, yields $\tilde{t}_{\rm coll}\approx$(2-5)$\times \mathrm{e}^{-s_{2}/2} \approx 0.09-0.21$,  consistent with the above estimate of $\tilde{t}_{\rm coll}$, which we finally estimate as $\tilde{t}_{\rm coll} = 0.2 \pm 0.1$.
The theory thus suggests that gravity has started dominating dense regions, corresponding to the onset of the first power law at $s=s_1$ only recently, i.e. for a short time $\tilde{t}_{\rm coll}$.  According to these determinations, the quiescent Polaris  region is quite young and has not even reached half its mean free-fall time yet.  Eventually, we expect it to start  forming detectable pre-stellar cores in a timescale of the order of its mean-free fall time, most likely in the  ``Saxophone" region, which entails most of the power-law part of its PDF \citep{schneider2013}. Taking $s_1 = s_{\rm G}$  yields an upper limit $\alpha_{\rm vir} \lesssim 1.2 $. 
The estimated $(b \mathcal{M})$ for Polaris yields mean Mach numbers $\mathcal{M} = 3.8^{+0.4}_{-0.4}$ and $\mathcal{M} = 2.5^{+0.3}_{-0.3}$ for  $b=1/3$ and $b=1/2$, respectively, consistent with the estimation of \citet{schneider2013}.

\section{Conclusion} \label{sec:conclu}

In this Letter, we have derived an analytical theory of the PDF of density fluctuations in supersonic turbulence in the presence of a gravity field in star-forming molecular clouds. The theory is based on a derivation of a combination of the coupled Navier-Stokes equations for the fluid motions and the Poisson equation for the gravity. The theory extends upon previous approaches \citep{pope1981,pope1985,pan2018,Pan2019A,Pan2019B} first by including gravity, second by considering the PDF as a dynamical system, not a stationary one. We derive rigorously the  transport equations  of the PDF, characterize its  evolution and determine the density threshold above which gravity strongly affects and eventually dominates the dynamics of the turbulence.  The theoretical results and diagnostics reproduce very well  numerical simulations of gravoturbulent collapsing clouds  (\S  \ref{sec:numerical}) and  various available observations (\S \ref{sec:obs}).

A major result of the theory is the characterization of two density regions in the PDF (see \S \ref{subsec:effectsofG}). A low density region where gravity does not affect \textit{significantly} the dynamics of turbulence so the PDF is the one of pure gravitationless turbulence, which resembles a lognormal form for isothermal, dominantly solenoidal turbulence. 
Then, above a density threshold, $s_{\rm G}$, given by Eqs.~(\ref{eq:gravitythresholdvir}) and  (\ref{eq:simulationsg}), gravity starts affecting \textit{significantly} the turbulence, essentially by increasing the velocity dispersion (thus the variance). Above this threshold,  $s>s_{\rm G}$, power-law tails develop over time  in the $s$-PDF, $f(s,t)\propto e^{-\alpha_s s}$, i.e. $p(\eta,t)\propto e^{-\alpha_\eta \eta}$  for the $\eta$-PDF of the surface density, as a direct consequence of the rising impact of gravity upon turbulence (see \S \ref{sec:evolutionPDF}). Within a typical timescale $\sim$$\tau_{\rm G}(s)=\tg\,\mathrm{e}^{-s/2}$, with $\tau_{\rm G,0} \equiv 1/\sqrt{4 \pi G \overline{\rho}}$, this yields the onset of a {\it first power law tail} with $\alpha_s \ge 2$, i.e. $\alpha_\eta \ge 4$. Later on, after a few $\tau_{\rm ff}(s)$ for a given density $s$, and/or at higher density, i.e. smaller scales, a {\it second power law} develops, with $\alpha_s=3/2$, i.e. $\alpha_\eta=2$. This is the signature of regions in free-fall collapse. 


Another important result of this study is to provide a procedure to relate the observed thresholds in {\it column density}, corresponding to the onset of the two power-law tails in the $\eta$-PDF, to the corresponding ones in {\it volume density} in the $s$-PDF (see \S\ref{sec:columndens} and App. A). Combined with the results of \S  \ref{subsec:effectsofG} and \S \ref{sec:evolutionPDF},  this allows to infer, \textit{from the observation of column-densities}, various physical parameters characterizing molecular clouds (or regions of), notably the virial parameter $\alpha_{\rm vir}$.  Moreover, the theory offers the possibility to date the clouds in units of $\tilde{t}_{\rm coll}$, i.e. {\it the time  since a statistically significant fraction of dense regions of the cloud started to collapse}, normalized to the cloud's mean free-fall time.  This explains why clouds exhibiting $\eta$-PDF with steep power laws ($\alpha_\eta \geq 3$) or extended lognormal parts are quiescent, since they have a short “age" $\tilde{t}_{\rm coll}$. This applies to  Polaris  \citep{andre2010,miville2010,schneider2013} (\S \ref{sec:obs}) but could also explain the quiescence of Chamaelon III \citep{deoliveira2014}.

The theory derived in this study allows the determination of the aforementioned volume and column density thresholds, $s_G,\eta_G$, and the characteristic timescales $\tau_{\rm G}(s),\tilde{t}_{\rm coll}$ (Eqns~(\ref{eq:gravitythresholdvir}),(\ref{eq:freefall}),(\ref{eq:transitoryfreefall}),(\ref{eq:linkedcritical}),(\ref{eq:simulationsg})). This yields {\it quantitative, predictive diagnostics}, from either simulations or observations, to determine precisely the relative impact of gravity upon  turbulence within star forming clouds/clumps and their evolutionary status. The theory thus provides a precise scale and clock to numericists and observers exploring star formation in MCs.
It provides a sound theoretical foundation and quantitative diagnostics to analyze observations or numerical simulations of star-forming regions and to characterize the evolution of the density PDF in various regions of MCs. This theoretical framework  provides a new vision on how gravitational collapse initiates and evolves within turbulent dense star-forming regions. 

\acknowledgments
The authors are grateful to Christoph Federrath  for sending us the PDFs of his simulations and to Nicola Schneider for the observational column densities presented in this article. We thank the anonymous referee for his/her insightful remarks that helped improving the manuscript.
We are also thankful  to  Benoit Commer\c con, J\'eremy Fensch, Guillaume Laibe  and Quentin Vigneron for helpful conversations. 

\appendix
\section{Derivation of the expression of the source terms Eqs 14-16} \label{ap:DetailsSgrav}

To obtain the source terms Eqs.~(\ref{eq:defSgrav}-\ref{eq:defSturb}) that appear in Eq.~(\ref{eq:dtdivu2ln}), we start by taking the divergence of Eq.~(\ref{eq::momentum})
\begin{eqnarray}
\partial_t  \left( \bm{\nabla} \cdot \bm{v} \right) + \bm{\nabla} \cdot  \left( \left\lbrace \bm{v} \cdot \bm{\nabla} \right\rbrace \bm{u} \right) + \bm{\nabla} \cdot  \left( \left\lbrace \bm{u} \cdot \bm{\nabla} \right\rbrace \bm{V} \right) +  \bm{\nabla} \cdot  \left( \left\lbrace \bm{V} \cdot \bm{\nabla} \right\rbrace \bm{V} \right) &=& - 4 \pi \, G \, \rho  - \bm{\nabla} \cdot  \left( \frac{1}{\rho} \bm{\nabla} P \right). \label{eq:divNS}
\end{eqnarray}
We then take the average of Eq.~(\ref{eq:divNS}) to obtain
\begin{eqnarray}
\partial_t  \left( \bm{\nabla} \cdot \bm{V} \right) +  \bm{\nabla} \cdot  \left( \left\lbrace \bm{V} \cdot \bm{\nabla} \right\rbrace \bm{V} \right) &=& - 4 \pi \, G \, \overline{\rho}, \label{eq:divNS}
\end{eqnarray}
where  $\partial_t  \left( \bm{\nabla} \cdot \bm{\overline{u}} \right) = \bm{\nabla} \cdot   \left( \overline{\left\lbrace \bm{v} \cdot \bm{\nabla} \right\rbrace \bm{u} } \right) = \bm{\nabla} \cdot  \left( \left\lbrace \overline{\bm{u}} \cdot \bm{\nabla} \right\rbrace \bm{V} \right) = \bm{\nabla} \cdot  \left(  \overline{\frac{1}{\rho} \bm{\nabla} P} \right)=0$, because the turbulent fields $\rho$ and $\bm{u}$ are statistically homogeneous and  because of the barotropic E.O.S $P=P(\rho)$. Then, by subtraction, we obtain
\begin{eqnarray}
\partial_t  \left( \bm{\nabla} \cdot \bm{u} \right) + \bm{\nabla} \cdot  \left( \left\lbrace \bm{v} \cdot \bm{\nabla} \right\rbrace \bm{u} \right) + \bm{\nabla} \cdot  \left( \left\lbrace \bm{u} \cdot \bm{\nabla} \right\rbrace \bm{V} \right)  &=& - 4 \pi \, G \, \overline{\rho} \left( \mathrm{e}^s - 1 \right)  - \bm{\nabla} \cdot  \left( \frac{1}{\rho} \bm{\nabla} P \right). 
\end{eqnarray}
We then note that $ \bm{\nabla} \cdot  \left( \left\lbrace \bm{u} \cdot \bm{\nabla} \right\rbrace \bm{V} \right) = \bm{u} \cdot \bm{\nabla} \left(\bm{\nabla} \cdot \bm{V} \right) + (\partial_i u_j) (\partial_j V_i) = (\partial_i u_j) (\partial_j V_i) =  \bm{\nabla} \bm{V} : \bm{\nabla} \bm{u}$, because $\rho$ is homogeneous, and by expanding $\bm{\nabla} \cdot  \left( \left\lbrace \bm{v} \cdot \bm{\nabla} \right\rbrace \bm{u} \right)$ we finaly obtain 
\begin{eqnarray}
\frac{\mathrm{D} \bm{\nabla} \cdot \bm{u}}{\mathrm{D} t} = - \bm{\nabla} \bm{u} : \bm{\nabla} \bm{u} - 2 \bm{\nabla} \bm{V} : \bm{\nabla} \bm{u} - 4 \pi \, G \, \overline{\rho} \left( \mathrm{e}^s - 1 \right) -  \bm{\nabla} \cdot  \left( \frac{1}{\rho} \bm{\nabla} P \right),
\end{eqnarray}
giving the expression of the source terms Eqs.~(\ref{eq:defSgrav}-\ref{eq:defSturb}).

\section{Reciprocal of Eqs 27-28.} \label{app:reciprocal}

In \S \ref{sec:evolutionPDF} we have shown that a conditional expectation $\left< \bm{\nabla}  \cdot \bm{u} | s, t \right> = h(t) \times \mathrm{e}^{a s}$, with $a>0$, would produce a $s$-PDF with a power law tail with exponent $\alpha_s=a+1$, i.e $f(s) \propto \mathrm{e}^{-(a+1)s}$ (Eqs.  \ref{eq:separ}-\ref{eq:solfexp}). We show here the reciprocal.

Let us assume that the $s$-PDF, $f$, is non-static and has a power-law tail with exponent $\alpha_s=a+1$ with $a>0$. More precisely, let us assume that 
\begin{equation}
f(s,t) = A(s,t)  \, \mathrm{e}^{-(a+1)s},
\end{equation} 
with a function $A(s,t)$ such as $A(s,t) \approx B(t) $ for $s \geq s_c$, for some $s_c$, where $B(t)$ is a $\mathcal{C}^1$ function of the time variable only, with a bounded derivative. Re-writing Eq. \ref{eq:evolvf} as 
\begin{equation}
 \left\lbrace \frac{\partial}{\partial s} + \frac{\partial \mathrm{ln} A}{\partial s} - a  \right\rbrace \left< \bm{\nabla}  \cdot \bm{u} | s, t \right> = \frac{\partial \mathrm{ln} A}{\partial t} ,
\end{equation}
one obtains
\begin{equation}
 \left< \bm{\nabla}  \cdot \bm{u} | s, t \right> = C(t) \times A(s,t)^{-1} \times \mathrm{e}^{a \, s} + A(s,t)^{-1} \,  \mathrm{e}^{as} \int_{s_i}^{s} \, \mathrm{e}^{-as'} \frac{\partial A}{\partial t}(s',t) \, \mathrm{d} s', \label{eq:reciprocal}
\end{equation}
with $C(t)$ a function of the time variable only and $s_i$ some fixed density. As $f$ is not stationary, $\left< \bm{\nabla}  \cdot \bm{u} | s, t \right>$ is not zero everywhere but, at any time $t$, there exists $s_0(t)$ such as $\left< \bm{\nabla}  \cdot \bm{u} | s_0(t), t \right> = 0$ to ensure $\overline{\bm{\nabla}  \cdot \bm{u}}=0$. Then we can fix the function $C(t)$ to write without any loss of generality:
\begin{equation}
\left< \bm{\nabla}  \cdot \bm{u} | s, t \right>  = A(s,t)^{-1} \, \mathrm{e}^{as} \int_{s_0(t)}^{s} \, \mathrm{e}^{-as'} \frac{\partial A}{\partial t}(s',t) \, \mathrm{d} s'. \label{eq:reciprocal}
\end{equation}
Then, because $a>0$ and $\partial_t B(t)$ is bounded,  the integral on the r.h.s of Eq. \ref{eq:reciprocal} is bounded and converges \textit{rapidly} towards $\int_{s_0(t)}^{+ \infty} \, \mathrm{e}^{-as'} \partial_t A(s',t)\, \mathrm{d} s' = I(t)$. The asymptotic behavior of $\left< \bm{\nabla}  \cdot \bm{u} | s, t \right>$ for large $s \geq s_c$  is thus
\begin{equation}
\left< \bm{\nabla}  \cdot \bm{u} | s, t \right> \approx I(t) \times B(t)^{-1} \times \mathrm{e}^{a \, s} = h(t) \times \mathrm{e}^{a \, s}.
\end{equation}

This shows that to a non-stationary $s$-PDF with a power-law tail of exponent $\alpha_s=a+1$ corresponds a conditional expectation $\left< \bm{\nabla}  \cdot \bm{u} | s, t \right> \approx h(t)  \times \mathrm{e}^{a s}$  for $s \gg 1$.

\section{Transitions to power-law tails} \label{app:columndens}

In \S \ref{sec:columndens} we derived a way to relate the volume density at which the $s$-PDF, $f(s)$, develop power-laws to the column density at which the $\eta$-PDF, $p(\eta)$, develops a similar behaviour. We call $s_{\rm crit}$ the  critical value corresponding to the beginning of a power-law tail in the $s$-PDF (Eq. (\ref{eq:linkedcritical}).  

Assuming ergodicity, we relate the volume fraction of regions with densities exceeding $s_{\rm crit}$ to the probability of finding a density exceeding $s_{\rm crit}$:
\begin{equation}
    \frac{V\left( s \geq s_{\rm crit} \right)}{V(\mathrm{cloud})} = \int_{s_{\rm crit}}^\infty f(s) \mathrm{d}s.
\end{equation}
We now want to evaluate the projected area of this volume onto the plane perpendicular to the line of sight $S( s \geq s_{\rm crit})$. Assuming statistical isotropy we get :
\begin{equation}
    \frac{S( s \geq s_{\rm crit})}{S(\mathrm{cloud})} \simeq \left( \frac{V\left( s \geq s_{\rm crit} \right)}{V(\mathrm{cloud})}\right)^{2/3}. \label{eq:surfacetovolume}
\end{equation}
We then identify regions on the observed area of the cloud contributing to the power-law in the $\eta$-PDF to regions included in the projected area $S( s \geq s_{\rm crit})$. This yields for the critical surface density $\eta_{\rm crit}$ at which the $\eta$-PDF transits to a power-law:
\begin{equation}
    \int_{\eta_{\rm crit}}^\infty p(\eta) \mathrm{d} \eta = \frac{S( s \geq s_{\rm crit})}{S(\mathrm{cloud})} \simeq \left( \int_{s_{\rm crit}}^\infty f(s) \mathrm{d}s \right)^{2/3}, 
\end{equation}
which is Eq. (\ref{eq:linkedcritical}).

\section{Numerical models} \label{app:numericalmodels}
In each simulation, gravity is switched on and sink particles are allowed to form after a state of fully developed turbulence has been reached, which determines the initial conditions at $t=t_0=0$ in the simulation. The associated transport equations for these simulations are:
\begin{eqnarray}
    \bm{V} &=& 0, \\
    \overline{\rho} &=& \rho_0, \\
     s&=&\mathrm{ln}\left(\rho/\rho_0\right), \\
    \frac{\mathrm{D} \bm{\nabla} \cdot \bm{u} }{\mathrm{D} t} &=& -\bm{\nabla} \bm{u} : \bm{\nabla} \bm{u} - 4 \pi G \rho_0 \left(\mathrm{e}^s  - 1 \right) \Theta(t) \nonumber \\
    &&- c_{\rm s}^2 \,  \bm{\nabla}^2 s + \bm{\nabla}  \cdot \bm{F}_{\rm stir},  \\
    \frac{\partial}{\partial t}\left[ \left< \left(\bm{\nabla}  \cdot \bm{u} \right)^n| s \right> f \right] &=& \left\lbrace 1 +\frac{\partial}{ \partial s}  \right\rbrace \left[\left< \left(\bm{\nabla}  \cdot \bm{u}\right)^{n+1} | s \right> f \right]  \nonumber \\
    && +  f \left<  \frac{\mathrm{D} \left(\bm{\nabla}  \cdot \bm{u}\right)^{n}}{\mathrm{D}t}| s \right>, \label{eq:transppdfsimu}
\end{eqnarray}
where $\rho_0$ is constant, $c_{\rm s} = 0.2$ km.$\mathrm{s}^{-1}$ is the sound speed, $\bm{\nabla}  \cdot \bm{F}_{\rm stir}$ is the divergence of the turbulent forcing, which is $0$ for a solenoidal driving, and $\Theta(t)$ is the Heaviside step function ensuring that gravity is plugged in at $t=0$. In all models the Mach number $\mathcal{M}$ slightly increases with time because of collapsing regions. For most models, this only amounts to a few percents except for the $\mathcal{M}\simeq 3$ simulations which start at $\mathcal{M}\simeq 2$ and end up at $\mathcal{M}\simeq 3-4$ because the virial parameter $\alpha_{\rm vir,0}= 5 \sigma_v ^2 /(6 G L_{\rm b}^2 \rho_{0})$ is very small (see their Table 1). We note that the aforementioned definition of  $\alpha_{\rm vir,0}$ taken from \citet{federrath2012,federrath2013} differs by the one we have introduced in \S \ref{subsec:effectsofG} by a factor $\pi/6 \simeq 1/2$, if the cloud size $L_{\rm c}$ is taken to be the box size $L_{\rm b}$.  As there is no unique way of translating the dimension of a cubic box into that of a spherical cloud and in order to simplify the comparison between the simulations and our calculations, we keep their notation and definition. 

Finally, to be consistent with the authors we describe the time evolution of the simulations by means of  the reduced time $\tilde{t} = t/\tff$,  which is the time in unit of mean free fall time $ \tff \equiv   \sqrt{\frac{3 \pi}{32 G \rho_0}}$, and by means of the star formation efficiency (SFE), which is set at $0 \%$ at the formation of the first sink particle. The authors only extracted the PDFs up to SFE$=20\%$ which we will thus refer to as the ‘‘long time" of the runs. 

\section{Model with one or two power-law tails.} \label{app:modelPLT}

In this section, we develop a simple model that allows to infer the global $s$-PDF of molecular clouds from the  observations of $\eta$-PDFs. We assume that the PDFs are simply continuous and have only one power-law at high densities and a lognormal cutoff at low densities: 
\begin{eqnarray}
  f(s) &=& A_1 \, \mathrm{e}^{-\frac{(s-\mu)^2}{2 \sigma_s^2}}, \, \,  s \leq s_{\rm crit} \nonumber \\
       &=& A_2 \, \mathrm{e}^{-\alpha_s \left(s-s_{\rm crit} \right)}, \, \, s \geq s_{\rm crit}.
\end{eqnarray}
Enforcing the continuity and normalisation of $f$ as well as the necessary condition $\overline{\mathrm{e}^s} =1$ (from our definition of $s$ in Eq.~(\ref{eq:defs})) we obtain:
\begin{eqnarray}
    A_1 &=& A_2 \, \mathrm{e}^{\frac{(s_{\rm crit}-\mu)^2}{2 \sigma_s^2}} \label{eq:obscontin}\\
    1 &=& \frac{1}{2}\,A_1 \sqrt{2 \pi \sigma_s^2}\,  \left[1 + \mathrm{erf}\left(\frac{s_{\rm crit} - \mu }{\sqrt{2}\sigma_s} \right) \right] + \frac{A_2}{\alpha_s}  \label{eq:obsnorm} \\
    1 &=& A_1 \sqrt{\frac{\pi}{2} \sigma_s^2}\, \mathrm{e}^{\mu+\frac{\sigma_s^2}{2}}  \left[1 + \mathrm{erf}\left(\frac{s_{\rm crit} - \mu -\sigma_s^2}{\sqrt{2}\sigma_s} \right) \right] \nonumber \\
    && + \frac{A_2 \, \mathrm{e}^{s_{\rm crit}}}{\alpha_s -1}. \label{eq:obsnorm2}
\end{eqnarray}
We now assume that the variance $\sigma_s$ in the lognormal part and the exponent $\alpha_s$ of the power-law tail are inferred from the observations of the $\eta$-PDF following \S\ref{sec:columndens}. More precisely, to obtain the variance $\sigma_s$, we use the formula of \citet{Burkhart2012}, $\sigma_\eta^2 = A_{\eta s} \times \sigma_s^2$, where $A_{\eta s}$  may depend on the forcing parameter $b$. For simulations of compressible turbulence without gravity and with solenoidal driving ($b=1/3$) they found from their best fit $A_{\eta s} \simeq 0.11$, while observations of molecular clouds yield a value $A_{\eta s} \simeq 0.12-0.16$ for a forcing parameter $b=0.5$, corresponding to a mixture of solenoidal and compressive driving. 
 From \S\ref{sec:columndens}, $A_2/\alpha_s$, thus $A_2$ is obtained from the observations. We are now left with a system of 3 equations for 3 unknown quantities, namely $s_{\rm crit}$, $\mu$ and $A_1$. We note that, in this model, the parameter $\mu$, that determines the peak of the lognormal part, is shifted to lower densities to ensure $\overline{\mathrm{e}^s} =1$. Injecting Eq.~(\ref{eq:obscontin}) into Eq.~(\ref{eq:obsnorm}) yields a closed equation for the variable $
x=\frac{s_{\rm crit} - \mu }{\sqrt{2}\sigma_s}$:
\begin{eqnarray}
    1 = A_2 \, \mathrm{e}^{x^2} \sqrt{2 \pi \sigma_s^2} \, \Phi(x) + \frac{A_2}{\alpha_s}, \label{eq:obsx}
\end{eqnarray}
with $\Phi(x) = \frac{1}{2} \left[1 + \mathrm{erf}(x) \right]$ the cumulative distribution function for the normal distribution. Eq.~(\ref{eq:obsnorm2}) is then used to obtain $\mu$ and then $s_{\rm crit}$.

In case where the $\eta$-PDF exhibits two power-law tails with exponent $\alpha_\eta =4$ and $\alpha_\eta =2$ we simply  assume the following functional form:
\begin{eqnarray}
  f(s) &=& A_1 \, \mathrm{e}^{-\frac{(s-\mu)^2}{2 \sigma_s^2}}, \, \,  s \leq s_1 \nonumber \\
       &=& A_2 \, \mathrm{e}^{-\alpha_1 \left(s-s_1\right)}, \, \, s_1\leq s \leq s_2\nonumber \\
       &=& A_2 \, \mathrm{e}^{-\alpha_1 \left(s_2-s_1\right)} \,\mathrm{e}^{-\alpha_2 \left(s-s_2 \right)}, \, \, s_2 \leq s,
\end{eqnarray}
with $\alpha_1=2$ and $\alpha_2=3/2$, and change the procedure as follows. First, we build the $s$-PDF as if there was only one power-law with exponent $\alpha_\eta =4$ in the $\eta$-PDF with the aforementioned procedure  to obtain $A_1$, $A_2$, $\mu$ and $s_1$.  We then use Eq.~(\ref{eq:linkedcritical}) to obtain $s_2$:
\begin{equation}
 \frac{A_2 \, \mathrm{e}^{-\alpha_1 \left(s_2-s_1\right)}}{\alpha_2} = \left(   \int_{\eta_{2}}^\infty p(\eta) \mathrm{d} \eta  \right)^{3/2},
\end{equation}
where $\eta_2$ is the column density at the beginning of the second power-law with exponent $\alpha_\eta =2$. This modified procedure, while simple to implement, is sufficiently accurate as $s_2$ is large and  thus regions with $s>s_2$ only represent a particularly small fraction of the total volume ($\lesssim 10^{-5}$).

We confront this procedure to observations in \S \ref{sec:obs}. Errors arising from the determination of $\eta_{\rm crit}$ and $\sigma_s$ from the observations yield an error  $\Delta s_{\rm crit}=\pm 0.3$ on $s_{\rm crit}$, which is reasonable.

\bibliography{PDF_evolution.bib}

\begin{thebibliography}{}
\expandafter\ifx\csname natexlab\endcsname\relax\def\natexlab#1{#1}\fi
\providecommand{\url}[1]{\href{#1}{#1}}
\providecommand{\dodoi}[1]{doi:~\href{http://doi.org/#1}{\nolinkurl{#1}}}
\providecommand{\doeprint}[1]{\href{http://ascl.net/#1}{\nolinkurl{http://ascl.net/#1}}}
\providecommand{\doarXiv}[1]{\href{https://arxiv.org/abs/#1}{\nolinkurl{https://arxiv.org/abs/#1}}}

\bibitem[{Andr{\'e} {et~al.}(2010)Andr{\'e}, Men'shchikov, Bontemps,
  K{\"o}nyves, Motte, Schneider, Didelon, Minier, Saraceno, Ward-Thompson,
  {et~al.}}]{andre2010}
Andr{\'e}, P., Men'shchikov, A., Bontemps, S., {et~al.} 2010, Astronomy \&
  Astrophysics, 518, L102

\bibitem[{Ballesteros-Paredes {et~al.}(2011)Ballesteros-Paredes,
  V{\'a}zquez-Semadeni, Gazol, Hartmann, Heitsch, \&
  Col{\'\i}n}]{ballesteros2011}
Ballesteros-Paredes, J., V{\'a}zquez-Semadeni, E., Gazol, A., {et~al.} 2011,
  Monthly Notices of the Royal Astronomical Society, 416, 1436

\bibitem[{Brunt {et~al.}(2010)Brunt, Federrath, \& Price}]{brunt2010}
Brunt, C.~M., Federrath, C., \& Price, D.~J. 2010, Monthly Notices of the Royal
  Astronomical Society, 403, 1507

\bibitem[{Burkhart \& Lazarian(2012)}]{Burkhart2012}
Burkhart, B., \& Lazarian, A. 2012, The Astrophysical Journal, 755, L19,
  \dodoi{10.1088/2041-8205/755/1/l19}

\bibitem[{Burkhart {et~al.}(2016)Burkhart, Stalpes, \& Collins}]{burkhart2016}
Burkhart, B., Stalpes, K., \& Collins, D.~C. 2016, The Astrophysical Journal
  Letters, 834, L1

\bibitem[{Cho \& Kim(2011)}]{cho2011}
Cho, W., \& Kim, J. 2011, Monthly Notices of the Royal Astronomical Society:
  Letters, 410, L8

\bibitem[{Collins {et~al.}(2012)Collins, Kritsuk, Padoan, Li, Xu, Ustyugov, \&
  Norman}]{collins2012}
Collins, D.~C., Kritsuk, A.~G., Padoan, P., {et~al.} 2012, The Astrophysical
  Journal, 750, 13

\bibitem[{De~Oliveira {et~al.}(2014)De~Oliveira, Schneider, Mer{\'\i}n, Prusti,
  Ribas, Cox, Vavrek, K{\"o}nyves, Arzoumanian, Puga,
  {et~al.}}]{deoliveira2014}
De~Oliveira, C.~A., Schneider, N., Mer{\'\i}n, B., {et~al.} 2014, Astronomy \&
  Astrophysics, 568, A98

\bibitem[{Donkov \& Stefanov(2018)}]{donkov2018}
Donkov, S., \& Stefanov, I.~Z. 2018, Monthly Notices of the Royal Astronomical
  Society, 474, 5588

\bibitem[{Federrath \& Klessen(2012)}]{federrath2012}
Federrath, C., \& Klessen, R.~S. 2012, \apj, 761, 156,
  \dodoi{10.1088/0004-637X/761/2/156}

\bibitem[{Federrath \& Klessen(2013)}]{federrath2013}
---. 2013, The Astrophysical Journal, 763, 51

\bibitem[{Federrath {et~al.}(2008)Federrath, Klessen, \&
  Schmidt}]{federrath2008}
Federrath, C., Klessen, R.~S., \& Schmidt, W. 2008, The Astrophysical Journal
  Letters, 688, L79

\bibitem[{Federrath {et~al.}(2010)Federrath, Roman-Duval, Klessen, Schmidt, \&
  Mac~Low}]{federrath2010}
Federrath, C., Roman-Duval, J., Klessen, R., Schmidt, W., \& Mac~Low, M.-M.
  2010, Astronomy \& Astrophysics, 512, A81

\bibitem[{Frisch(1995)}]{frisch1995}
Frisch, U. 1995, Turbulence: the legacy of AN Kolmogorov (Cambridge university
  press)

\bibitem[{Girichidis {et~al.}(2014)Girichidis, Konstandin, Whitworth, \&
  Klessen}]{girichidis2014}
Girichidis, P., Konstandin, L., Whitworth, A.~P., \& Klessen, R.~S. 2014, The
  Astrophysical Journal, 781, 91

\bibitem[{Guszejnov {et~al.}(2018)Guszejnov, Hopkins, \&
  Grudić}]{Guszejnov2018}
Guszejnov, D., Hopkins, P.~F., \& Grudić, M.~Y. 2018, Monthly Notices of the
  Royal Astronomical Society, 477, 5139, \dodoi{10.1093/mnras/sty920}

\bibitem[{Kainulainen {et~al.}(2009)Kainulainen, Beuther, Henning, \&
  Plume}]{kainulainen2009}
Kainulainen, J., Beuther, H., Henning, T., \& Plume, R. 2009, Astronomy \&
  Astrophysics, 508, L35

\bibitem[{Kainulainen {et~al.}(2006)Kainulainen, Lehtinen, \&
  Harju}]{kainulainen2006}
Kainulainen, J., Lehtinen, K., \& Harju, J. 2006, Astronomy \& Astrophysics,
  447, 597

\bibitem[{Klessen(2000)}]{klessen2000}
Klessen, R.~S. 2000, The Astrophysical Journal, 535, 869

\bibitem[{Kritsuk {et~al.}(2007)Kritsuk, Norman, Padoan, \&
  Wagner}]{kritsuk2007}
Kritsuk, A.~G., Norman, M.~L., Padoan, P., \& Wagner, R. 2007, The
  Astrophysical Journal, 665, 416

\bibitem[{Kritsuk {et~al.}(2010)Kritsuk, Norman, \& Wagner}]{kritsuk2010}
Kritsuk, A.~G., Norman, M.~L., \& Wagner, R. 2010, The Astrophysical Journal
  Letters, 727, L20

\bibitem[{Ledoux \& Walraven(1958)}]{ledoux1958}
Ledoux, P., \& Walraven, T. 1958, in Astrophysics II: Stellar
  Structure/Astrophysik II: Sternaufbau (Springer), 353--604

\bibitem[{Lee {et~al.}(2015)Lee, Chang, \& Murray}]{lee2015}
Lee, E.~J., Chang, P., \& Murray, N. 2015, The Astrophysical Journal, 800, 49

\bibitem[{Lemaster \& Stone(2008)}]{lemaster2008}
Lemaster, M.~N., \& Stone, J.~M. 2008, The Astrophysical Journal Letters, 682,
  L97

\bibitem[{Miville-Desch{\^e}nes {et~al.}(2010)Miville-Desch{\^e}nes, Martin,
  Abergel, Bernard, Boulanger, Lagache, Anderson, Andr{\'e}, Arab, Baluteau,
  {et~al.}}]{miville2010}
Miville-Desch{\^e}nes, M.-A., Martin, P., Abergel, A., {et~al.} 2010, Astronomy
  \& Astrophysics, 518, L104

\bibitem[{Miville-Desch{\^e}nes {et~al.}(2017)Miville-Desch{\^e}nes,
  Salom{\'e}, Martin, Joncas, Blagrave, Dassas, Abergel, Beelen, Boulanger,
  Lagache, {et~al.}}]{miville2017}
Miville-Desch{\^e}nes, M.-A., Salom{\'e}, Q., Martin, P., {et~al.} 2017,
  Astronomy \& Astrophysics, 599, A109

\bibitem[{Orkisz {et~al.}(2017)Orkisz, Pety, Gerin, Bron, Guzm{\'a}n, Bardeau,
  Goicoechea, Gratier, Le~Petit, Levrier, {et~al.}}]{orkisz2017}
Orkisz, J.~H., Pety, J., Gerin, M., {et~al.} 2017, Astronomy \& Astrophysics,
  599, A99

\bibitem[{Pan {et~al.}(2018)Pan, Padoan, \& Nordlund}]{pan2018}
Pan, L., Padoan, P., \& Nordlund, {\AA}. 2018, The Astrophysical Journal
  Letters, 866, L17

\bibitem[{Pan {et~al.}(2019{\natexlab{a}})Pan, Padoan, \& Nordlund}]{Pan2019B}
---. 2019{\natexlab{a}}, The Astrophysical Journal, 881, 155

\bibitem[{Pan {et~al.}(2019{\natexlab{b}})Pan, Padoan, \& Nordlund}]{Pan2019A}
---. 2019{\natexlab{b}}, The Astrophysical Journal, 876, 90

\bibitem[{Passot \& V{\'a}zquez-Semadeni(1998)}]{passot1998}
Passot, T., \& V{\'a}zquez-Semadeni, E. 1998, Physical Review E, 58, 4501

\bibitem[{Pope(1981)}]{pope1981}
Pope, S. 1981, The Physics of Fluids, 24, 588

\bibitem[{Pope \& Ching(1993)}]{pope1993}
Pope, S., \& Ching, E.~S. 1993, Physics of Fluids A: Fluid Dynamics, 5, 1529

\bibitem[{Pope(1985)}]{pope1985}
Pope, S.~B. 1985, Progress in energy and combustion science, 11, 119

\bibitem[{Schneider {et~al.}(2013)Schneider, Andr{\'e}, K{\"o}nyves, Bontemps,
  Motte, Federrath, Ward-Thompson, Arzoumanian, Benedettini, Bressert,
  {et~al.}}]{schneider2013}
Schneider, N., Andr{\'e}, P., K{\"o}nyves, V., {et~al.} 2013, The Astrophysical
  Journal Letters, 766, L17

\bibitem[{Schneider {et~al.}(2012)Schneider, Csengeri, Hennemann, Motte,
  Didelon, Federrath, Bontemps, Di~Francesco, Arzoumanian, Minier,
  {et~al.}}]{schneider2012}
Schneider, N. e.~a., Csengeri, T., Hennemann, M., {et~al.} 2012, Astronomy \&
  Astrophysics, 540, L11

\bibitem[{Truelove {et~al.}(1997)Truelove, Klein, McKee, Holliman~II, Howell,
  \& Greenough}]{truelove1997}
Truelove, J.~K., Klein, R.~I., McKee, C.~F., {et~al.} 1997, The Astrophysical
  Journal Letters, 489, L179

\bibitem[{Vazquez-Semadeni(1994)}]{vazquez1994}
Vazquez-Semadeni, E. 1994, The Astrophysical Journal, 423, 681

\bibitem[{Vazquez-Semadeni \& Garcia(2001)}]{Vazquez_Semadeni_2001}
Vazquez-Semadeni, E., \& Garcia, N. 2001, The Astrophysical Journal, 557, 727,
  \dodoi{10.1086/321688}

\end{thebibliography}
\bibliographystyle{aasjournal}



\end{document}